\documentclass[a4paper,fleqn]{cas-dc}

\usepackage[numbers]{natbib}

\def\tsc#1{\csdef{#1}{\textsc{\lowercase{#1}}\xspace}}
\tsc{WGM}
\tsc{QE}
\tsc{EP}
\tsc{PMS}
\tsc{BEC}
\tsc{DE}

\begin{document}
\let\WriteBookmarks\relax
\def\floatpagepagefraction{1}
\def\textpagefraction{.001}

\shorttitle{Antiferromagnetism in a nanocrystalline high entropy oxide (Co,Cu,Mg,Ni,Zn)O}
\shortauthors{Nandhini J Ushrani et~al.}

\title [mode = title]{Antiferromagnetism in a nanocrystalline high entropy oxide (Co,Cu,Mg,Ni,Zn)O : Magnetic constituents and surface anisotropy leading to lattice distortion} 
\author[1]{Nandhini J. Usharani}
\credit{Conceptualization, Methodology, Investigation, and Writing - Original Draft}
\address[1]{Nano Functional Materials Technology centre, (NFMTC), Department of Metallurgical and Materials engineering, Indian Institute of Technology Madras, Chennai-600036, India.}
\author[1]{Anikesh Bhandarkar}
\credit{Investigation}
\author[2]{Sankaran Subramanian}
\credit{Validation and Writing - Review and Editing}
\address[2]{Sophisticated Analytical Instrument Facility, Indian Institute of Technology Madras, Chennai- 600036, India.}
\author[1]{Subramshu S. Bhattacharya}[orcid=0000-0002-6865-0822]
\credit{Conceptualization, Methodology, Supervision, Project administration, Validation, Writing - Review and Editing}
\cormark[1]
\ead{ssb@iitm.ac.in}
\cortext[cor1]{Corresponding author}

\begin{abstract}
For the first time, this study shows that distortion in a crystal structure due to magnetic effect is possible in a lattice with extreme chemical disorder. The multicomponent equimolar–transition metal oxide (ME-TMO), (Co,Cu,Mg,Ni,Zn)O, which is a high entropy oxide, has been attracting a lot of attention due to its unique application potential in many fields including electrochemical energy storage. In the present investigation, nanocrystalline ME-TMO was synthesised by three bottom-up methods. The presence of distortion in the rocksalt crystal structure, revealed by X-ray diffraction and Raman spectroscopy, and correlated with magnetic measurements from SQUID and EPR studies could be attributed to the additive effects of exchange striction (from the magnetic constituents) and magnetic anisotropy (from the decreased crystallite size). For the first time, iron has been doped into ME-TMO, to show that a higher amount of magnetic constituent increases the distortion in the lattice. Nanocrystalline ME-TMO also showed a “core-shell” magnetic behavior below the bifurcation temperature arising from the uncompensated or canted spin at the surface. Neel temperature of the nanocrystalline ME-TMO is reported for first time to be as high as 700 K. This study helps unravel the structure and magnetic properties of such high entropy materials, and augurs a definite scope for better understanding of the factors influencing the crystal structure in high entropy oxides.
\end{abstract}

\begin{keywords}
High entropy oxide \sep Distortion \sep X-ray diffraction (XRD) \sep Magnetic properties 
\end{keywords}
\maketitle
\section{Introduction}
The (Co,Cu,Mg,Ni,Zn)O multicomponent equimolar transition metal oxide (ME-TMO) has been garnering a lot of interest in the past few years ever since Rost \textit{et al.}\cite{Rost2015}  reported, for the first time, the formation of a phase-pure rocksalt structure that was stabilised by entropy. Thereafter, such transition metal based “high entropy oxides” (HEO) having 5 or more cations have attracted the research community due to the prospect of synergistic functional properties such as high dielectric constant of 3000 with low loss at room temperature and as an excellent cathode material for Li-ion batteries with high specific capacity and stability of more than 500 cycles.\cite{Berardan2016, Sarkar2018} However, in all these cases the crystal structure of the ME-TMO has been reported to be of the (fm$\overline{3}$m) rocksalt structure with minor deviations being explained by the Jahn-Teller effect.\cite{Berardan2017} In this investigation, an attempt was made to understand and explore the other possible factors that would result in a phase-pure distorted rocksalt (or monoclinic) structure in the ME-TMO by systematic characterisation using different structural and other analytical tools.  

In an ME-TMO system there could be at least three factors that cause a lowering of symmetry in the crystal structure. For example, antiferromagnetic ordering in the lattice has been reported to cause distortion in the rocksalt lattice as in the case of intrinsic oxides of NiO and CoO.\cite{Smart1951} Hence, it is speculated that similar factors could cause distortion in the present ME-TMO structure due to the presence of Ni and Co. This kind of geometric frustration due to antiferromagnetism has also been observed in other rocksalt structures such as Li$_{3}$Mg$_{2}$RuO$_{6}$.\cite{Derakhshan2008} Jahn-Teller distortion in the system would be a second factor.\cite{Dalverny2010} Finally, the presence of multiple ions with differing bond lengths would also cause distortion in the structure. 

Some of the principal constituents such as NiO and CoO have a pseudo-cubic structure with a little distortion below its Neel temperature.\cite{Roth.W.L1958,Lee2016,Li1955,Rooksby1948} However, above the Neel temperature, the distorted cubic structure system stabilises into an NaCl (rocksalt) lattice arrangement. NiO and CoO exhibit type II antiferromagnetic structures due to the ordering of spin states, as described by Roth \textit{et al.}\cite{Roth.W.L1958}   The lowering of symmetry in the antiferromagnetic state is caused by two factors: exchange striction and magnetostriction.\cite{Kanamori1957} The trigonal distortion in the structure of systems such as NiO and CoO in the antiferromagnetic (AFM) state has been well studied.\cite{Smart1951, Roth.W.L1958, Kanamori1957} Exchange striction is change in the volume of the unit cell resulting from a lowering of the exchange interaction energy by an increase in the interatomic distance.\cite{Greenwald1950} Magnetostriction, on the other hand, is the volume change caused due to the magnetic anisotropic energy and depends on the orientation of the magnetic moment. Bean \textit{et al.}\cite{Bean1962} developed a theoretical relationship for the free energy of a magnetic material based on molecular field approximation, which shows that change in the volume due to the exchange interaction increase reduces the free energy of the structure and Bartell \textit{et al.}\cite{Bartel1970} have shown that the equilibrium change in angle and lattice parameter is proportional to the J interaction, and the free energy is reduced when there is change in the volume of the structure. From these it can be concluded that when the J integral is high, the lattice parameters and angular deviation from the cubic structure is high. In this work, the structure of (Co,Cu,Mg,Ni,Zn)O was investigated to find the distortion due to exchange and magnetostriction through analysis of X-ray diffraction data and spectroscopic measurements, and correlated to the magnetic property through magnetic measurements. Distortion in the structure due to magnetism is expected to decrease with the crystallite size, as the exchange interaction decreases which can also be observed as a decrease in the Neel temperature.\cite{Rinaldi-montes2016} However, in a few cases, the distortion has been observed to increase with decrease in the crystallite size.\cite{Kremenovic2012,Golosovsky2001} In this work, at an attempt has been made to explore and develop the mechanism of distortion in the ME-TMO (Co,Cu,Mg,Ni,Zn)O high entropy oxide. 

\section{Materials and Methods}

\subsection{Experimental materials and synthesis process}
Nickel (II) nitrate hexahydrate (99 \%, Alfa Aesar), cobalt (II) nitrate hexahydrate (99 \%, Alfa Aesar), magnesium nitrate hexahydrate (99 \%, Alfa Aesar), copper nitrate pentahydrate (99 \%, Alfa Aesar), zinc nitrate hexahydrate (99 \%, Alfa Aesar), iron nitrate nonahydrate (99 \%, Alfa Aesar) were dissolved in de-ionised water to prepare the precursor solution at one molar concentration.  The same precursor was used for all the three synthesis processes (RCP, FSP and NSP). A brief description of the synthesis processes are given subsequently, while the other details are available elsewhere.\cite{Sarkar2017}

In the FSP process, an ultrasonic nebuliser was used to convert the precursor solution into a fine aerosol mist and fed at the rate of 1 ml/min to a self-sustaining flame using oxygen as a carrier gas. The flame was generated using a diffusion burner in a chamber by using liquified petroleum gas as the fuel and O$_{2}$ as the combustion gas. The flow rates of the gases were 30 sccm (standard cubic centimetres per minute) of LPG and 15 slm (standard litres per minute) of oxygen. The nebulised aerosol pyrolysed in the flame and formed oxide by homogenous nucleation. Growth of the oxide particles was suppressed due to the short residence time in the flame and the particle quenched to room temperature because of the high flow rate in the flame chamber maintained at a subambient pressure of 60 mm Hg by a pumping system. The particles were collected by a suitable powder collector unit. In the NSP process, the precursor solution was nebulised under similar conditions as in the FSP and the aerosol mist sent into a tubular furnace maintained at 1423 K. The NSP set-up was maintained at a pressure of 60 mm Hg and powders collected by a filtering system. In the RCP process, the precursor was added drop-wise to ammonia solution maintained at a pH of 11, which was well above the saturation pH of the individual constituent oxides, forming a hydroxylated precipitate. The filtered precipitate was dehydrated in an air oven at 393 K for 8 hours and calcined in a box furnace at a 1173 K for 30 mins. 

\subsection{Characterization}
The synthesised powders were characterised using a powder X-ray diffractometer (Panalytical) fitted with a monochromatic copper source operating at a current and voltage setting of 30 mA and 45 kV respectively with a collection rate of 0.02$^{\circ}$/20 s. Rietveld refinement was carried out using the Fullprof software suite\textsuperscript{\textregistered}. Peaks were fitted with Pseudo-Voigt function. Background method used was linear interpolation of a set of data points. The refined parameters were scale factor, background, lattice constants, FWHM and shape parameters. Crystallite size was calculated from the X-ray diffraction pattern using the Scherrer formula and the instrumental broadening was subtracted by using a standard Si scan.  Magnetic measurements were made using a SQUID magnetometer (Quantum Design) from 5 K to 300 K at a constant field of 500 Oe and M-H measurements were taken from -70000 Oe to 70000 Oe at 10 K and at 300 K while a vibrating sample magnetometer, VSM (Lakeshore, 7460) was used for high temperature magnetic measurements (from room temperature to 900 K). Raman spectroscopy (Horiba Jobin Yvon HR 800) was carried out using a 633 nm He–Ne laser in the range of 200–1800 cm$^{-1}$. All the spectra were the result of 20 accumulations with each lasting for 20 s. EPR (JEOL, Model JES FA200) was done in the X band frequency at room temperature. 

\section{\label{sec:resultsanddiscussion}Results and discussion}

\subsection{\label{sec:structure}Structural analysis}

\begin{figure}
    \centering
    \includegraphics[width=8.6cm,height=7.5cm]{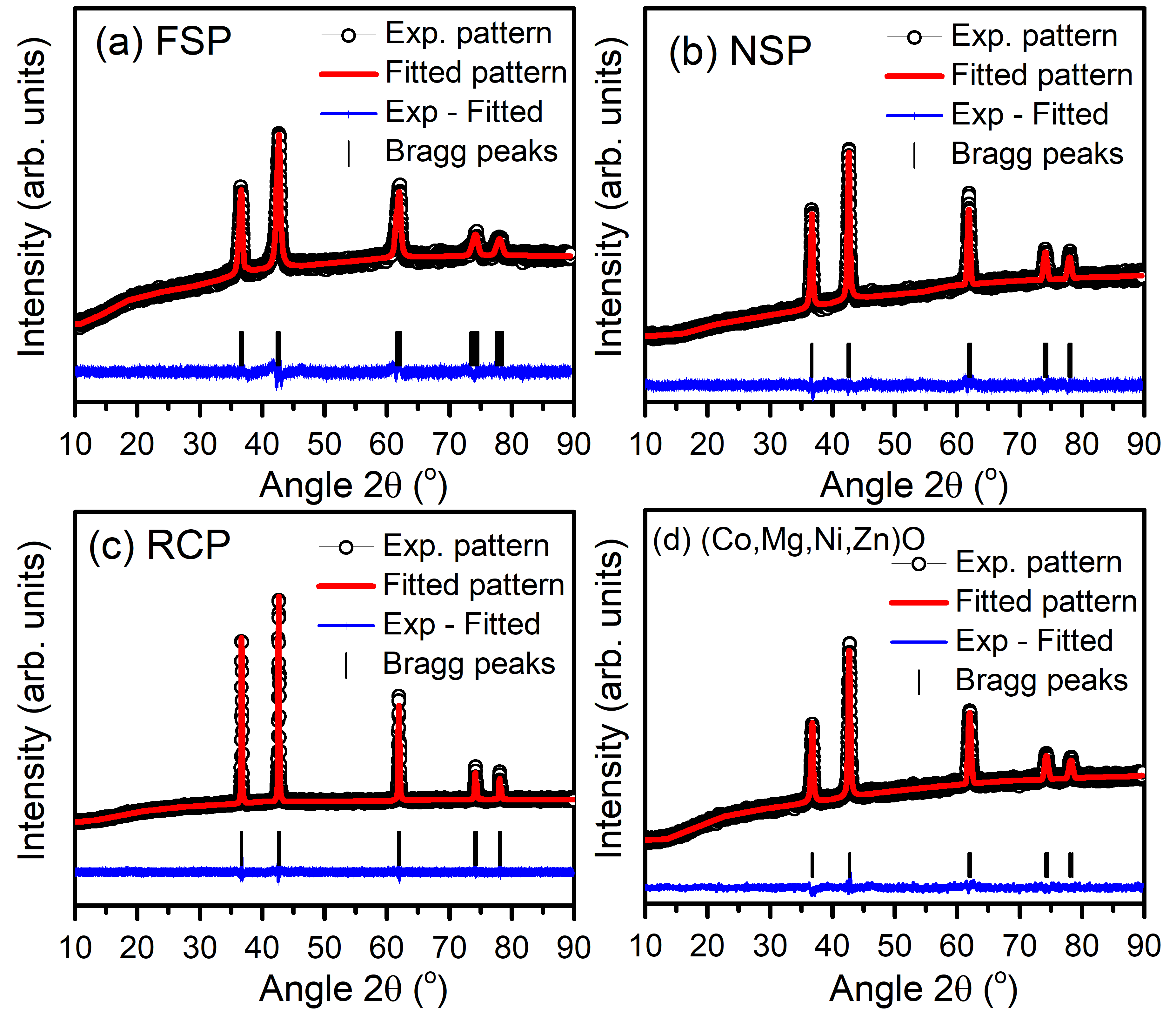}
    \caption{XRD patterns and Rietveld refinement fits of (Co,Cu,Mg,Ni,Zn)O powders synthesised by (a) FSP, (b) NSP, (c) RCP and (d) (Co,Mg,Ni,Zn)O synthesised through FSP (hkl planes are given in the supplementary table S1)}
    \label{fig:XRD}
\end{figure}

Till date, the crystal structure of the multicomponent system (Co,Cu,Mg,Ni,Zn)O has been reported to be of the rocksalt type (see, for example, Ref. 1,2,19). In the present study, the X-ray diffraction patterns of the powders synthesised by all three processes (FSP, NSP and RCP) still exhibit the characteristic fingerprint peak positions of rocksalt/NaCl structure as shown in figure  \ref{fig:XRD}. Careful observation revealed a minor asymmetry in all the peaks regardless of the synthesis process used. However, the degree of asymmetry varied with the process used for synthesis and was observed to be the most prominent in the FSP process (which had the least residence time) and the least in the RCP powders. 
Originally, Sarkar  \textit{et al.}\cite{Sarkar2017} had attributed the asymmetry to a second rocksalt phase in case of the FSP process, which was claimed to disappear on calcination at 1273 K for one hour. But, on closer inspection it was seen that the asymmetry had only reduced and not disappeared. By carrying out Rietveld refinement of the X-ray diffraction patterns of (Co,Cu,Mg,Ni,Zn)O, the asymmetry in the peaks was deconvoluted. The asymmetry in the peak was a resultant of peaks with nearby d-spacing and the well-matched simulated pattern showed that the structure belonged to the space group (C 1 2/m 1) as observed for the monoclinic phase of NiO and CoO.\cite{Lee2016} The lattice parameters and the Rietveld fit values are listed in table  \ref{table:xrd}.

\begin{table*}[width=.9\textwidth,cols=9,pos=h]
\caption{\label{table:xrd}Monoclinic lattice parameters obtained from Rietveld refinement of XRD data and crystallite size of ME-TMO}
\begin{tabular}{p{3.2cm}ccccp{2.3cm}ccp{2.8cm}}
\toprule
 &\multicolumn{4}{c}{\centering Monoclinic lattice parameters}&&&&\\
 \cline{2-5}
 Process &\centering{a (\AA)}&\centering b (\AA)&\centering c (\AA)&\centering {$\beta(^{\circ})$}&\centering  Cell volume (\AA$^{3}$)& Distortion (\%)&Chi$^{2}$&Crystallite size (nm) \\ \midrule
 RCP&5.191&2.993&2.996&\centering{125.206}&\centering38.070&0.171&2.06&81\\
 NSP&5.191&2.998&2.992&\centering{124.929}&\centering38.191&0.488&1.52&25\\
FSP&5.192&2.994&2.997&\centering{124.670}&\centering38.356&0.916&1.70&15\\
Ideal cubic structure\textsuperscript{a}&5.188&2.995&2.995&\centering{125.264}&\centering38.030&-&-&-\\
(Co,Mg,Ni,Zn)O -FSP\textsuperscript{b}&5.182&2.993&2.996&124.902&\centering38.109&0.626&1.58&33\\
\bottomrule
\end{tabular}
    
         \textsuperscript{a} Cubic lattice parameter of (Co,Cu,Mg,Ni,Zn)O used to obtain the monoclinic lattice parameters is a=4.236 \AA \\
         \textsuperscript{b} Cubic lattice parameter of (Co,Mg,Ni,Zn)O used to obtain the monoclinic lattice parameters is a=4.231 \AA \\
    
\end{table*}

\begin{figure}
    \centering
    \includegraphics[width=5.2cm,height=7.5cm]{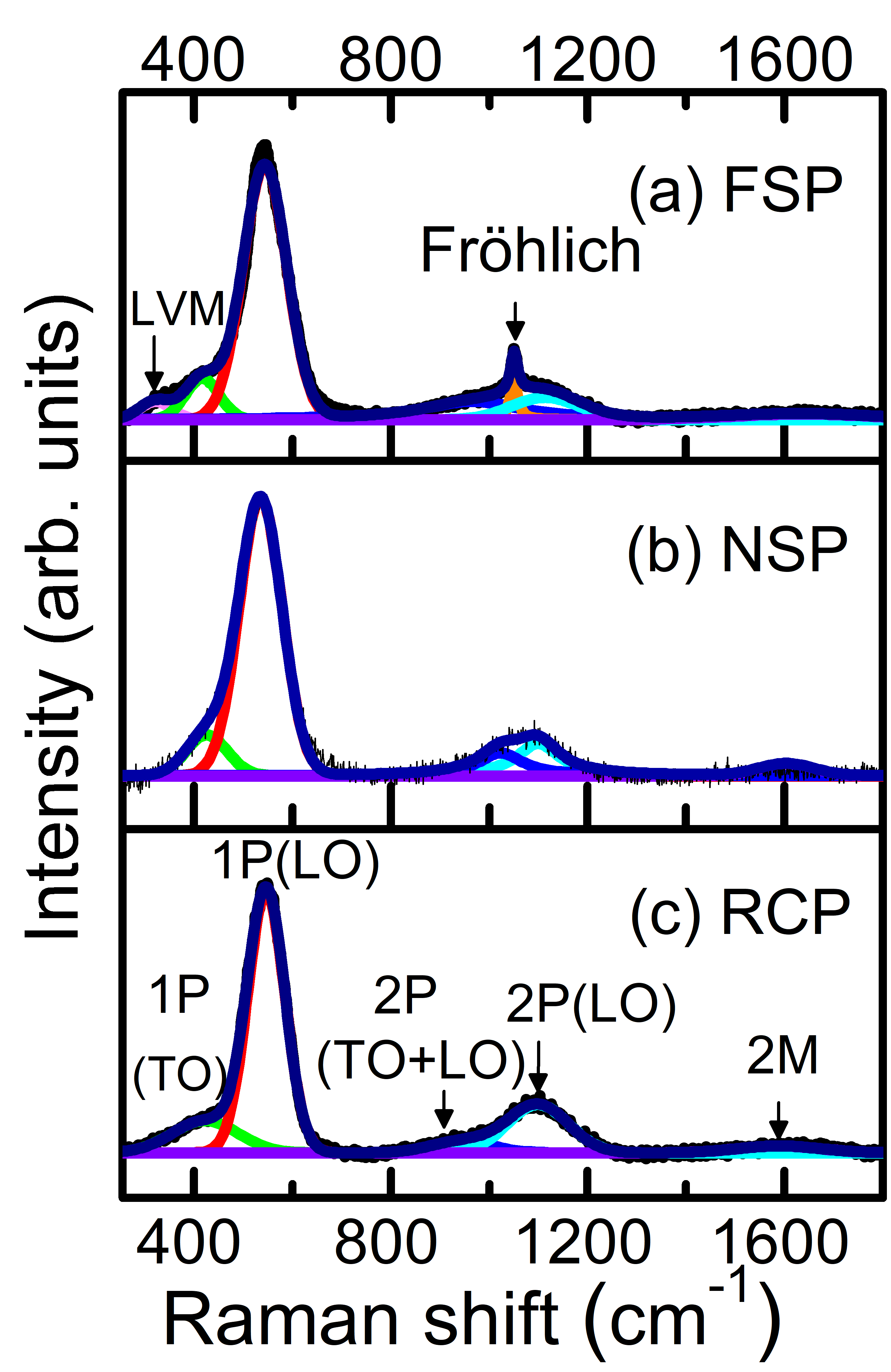}
    \caption{Raman spectra of (Co,Cu,Mg,Ni,Zn)O powders synthesised by (a) FSP, (b) NSP and (c) RCP}
    \label{fig:Raman}
\end{figure}

Any rocksalt cubic lattice can also be described in terms of an equivalent monoclinic structure. If the asymmetry in the peaks were to be ignored and the (Co,Cu,Mg,Ni,Zn)O system considered as a pure rocksalt structure, the cubic lattice constant works out to 4.23 {\AA}, while the equivalent monoclinic lattice parameters, a should have been 5.18 {\AA} with b and c each being 2.99 {\AA}  and angle $\beta$ = 125.264$^{\circ}$. (Lattice parameter of a cubic lattice can be transformed to the monoclinic lattice parameters using trigonometric functions \cite{Lee2016}). Any deviation in these values is a clear indication of distortion from the cubic lattice structure. The lattice parameters listed in table \ref{table:xrd}, show the deviation in the transformed monoclinic lattice parameters indicating a distortion in the structure. It is observed that the volume of the unit cell was the highest in case of the FSP process and lowest in case of RCP, showing higher distortion in the FSP powders.
The distortion in the structure can be also observed from the Raman spectra shown in figure \ref{fig:Raman}. The one phonon (1P) mode is absent in an ideal cubic structure and can be seen only in the presence of defects or distortion in the structure (as observed in NiO and CoO). The peak near 420 cm$^{-1}$ was assigned to 1P TO mode.  It can be noticed that the peak at 367 cm$^{-1}$ was present only in the FSP sample, which could be due to local vibration mode (LVM) caused by surface states.\cite{Duan2012}  The peak at 935 cm$^{-1}$ belonged to the 2P TO+LO vibration mode. The peak at 1090 cm$^{-1}$ was attributed to the 2P LO mode. The assigned vibrations were based on a structure with similar symmetry to the ME-TMO such as NiO or CoO.\cite{Usharani2019,Mironova-ulmane2008,Mironova-Ulmane2007}  The peak near 1600 cm$^{-1}$ was the magnon mode, which hinted that the material was antiferromagnetic (AFM) in nature.\cite{Chou1976}
Since it is well-established that the 1P LO mode of vibration observed during Raman spectroscopy is sensitive to the defect concentration in the material, the LO mode was used in the uncertainty relation to quantify the phonon lifetime from  Eq.~(\ref{eq:phonon})
\begin{equation}
\left\{
\frac{\Delta x}{h}= \frac{1}{t}%
\right\}.
\label{eq:phonon}
\end{equation}
where t is the phonon lifetime, h is Planck’s constant = 5.3 x 10$^{-12}$ cm$^{-1}$s, and $\Delta$x is the full width at half maximum (FWHM) of the 1P LO mode in the Raman spectrum.\cite{Bergman1999} The FSP powders had the shortest lifetime of 0.050 ps, while the NSP powders had 0.051 ps and the RCP powders 0.059 ps. As the lifetime of the phonon decreases with increasing defect concentration in the material, the FSP powders had evidently the highest defect concentration while the RCP powders had the least. This was attributed to the fact that the residence time during RCP is the highest, thereby allowing the material to form the most equilibrium structure with minimal defect states.\\
The influence of antiferromagnetic (AFM) ordering was analysed from magnetometry studies. The $\chi$-T plots of ME-TMO synthesised by all the three processes showed antiferromagnetic behaviour at room temperature. Magnetic interactions in a system cannot be explained in terms of the rule of mixtures, as the interaction of an unpaired electron of a cation with the empty orbital of a diamagnetic cation would result in a ferromagnetic exchange interaction. This has been observed in case of dilute magnetic semiconductors (DMS) where low dopant concentrations of transition metal oxides (TMO) in non-magnetic oxides like ZnO and MgO causes ferromagnetic interaction,\cite{Chanda2017,Ali2019} but when the dopant concentration is more than the percolation limit, the distances between the dopant ions decrease and the interaction becomes antiferromagnetic. All the possible dopant and oxide combinations present in the system show antiferromagnetic behaviour beyond the percolation limit.\cite{Xu2010,Theyvaraju2015,Coey2005} Since 3 out of 5 (60 \%) ions in the cationic sublattice of the ME-TMO are magnetic, an AFM interaction can be expected. The reason for the lower “monoclinic” symmetry could then be possibly explained by exchange striction caused by spin alignment in the antiferromagnetic lattice of NiO or CoO lattice planes.\cite{Lee2016, Bartel1971} Here, the spins get ferromagnetically aligned in one plane and oppositely aligned in the neighbouring plane. To reduce the interaction energy, the plane with the parallel spins expands and the neighbouring plane with the oppositely aligned spins contracts resulting in distortion in the structure. In case of NiO the plane is [111] while it is [100] for CoO.\cite{Roth.W.L1958, Kanamori1957} The actual magnetic anisotropy directions of NiO and CoO would play a relatively insignificant role in the direction of the distortion in the ME-TMO lattice as the cations are distributed uniformly, and so, the effect of magnetic anisotropy would be different from that in the intrinsic (unary) oxide. However, studies on neutron diffraction of micrograin ME-TMO have reported Jahn-Teller distortions due to the copper ions, but no magnetic distortion.\cite{Jimenez-segura2019, Zhang2019} If there is to be exchange striction in the lattice, exchange interaction (J) should be high for the lattice with greater distortion and will be reflected in terms of a higher magnetisation (as the two are directly proportional).\\

\begin{figure*}
\centering
\includegraphics[width=14.6cm,height=4cm]{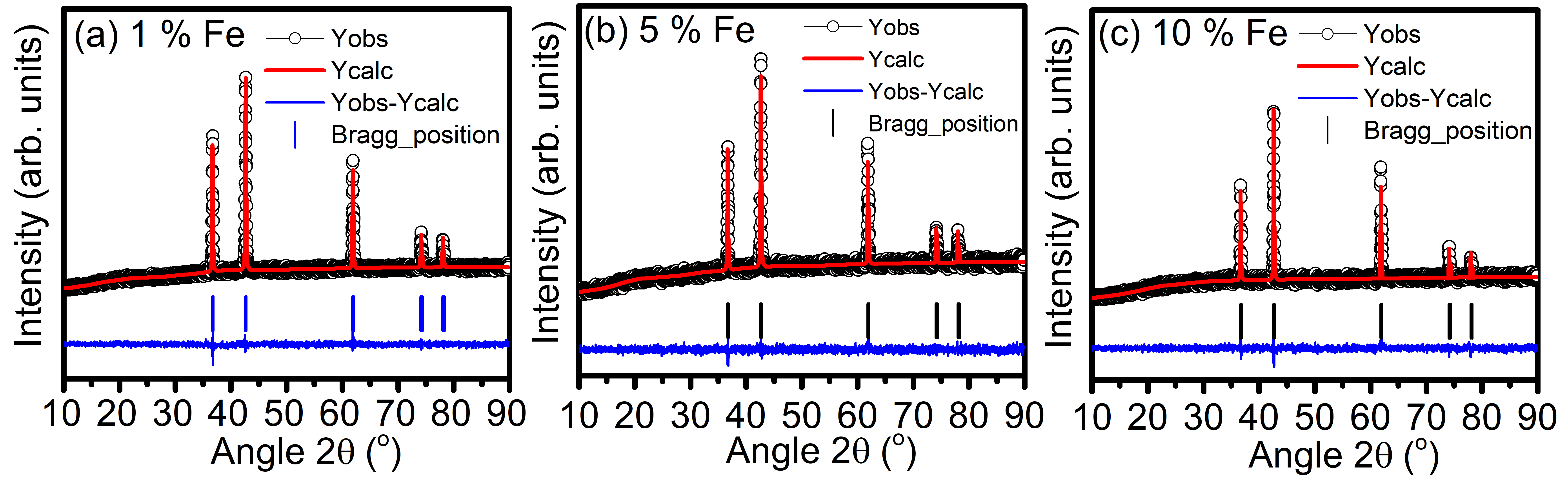} 
\caption{\label{fig:xrd2}X-ray diffraction patterns with Rietveld refinement fits of the (Co,Cu,Mg,Ni,Zn)O powders doped with iron (a) 1 \%, (b) 5 \% and (c) 10 \% (hkl planes are given in the supplementary table S2) }
\end{figure*}

\begin{figure}
    \centering
    \includegraphics[width=6.6cm,height=5.5cm]{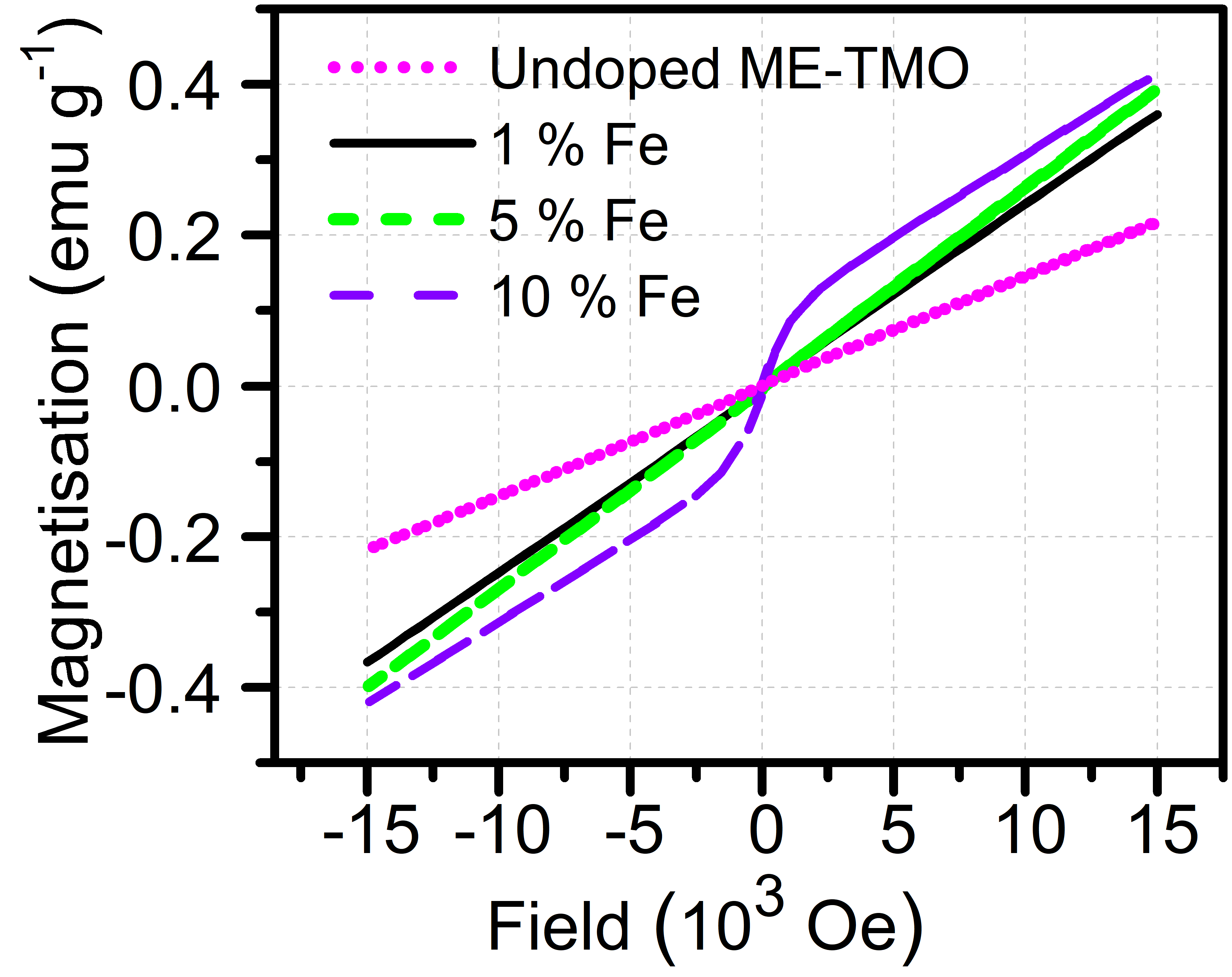}
    \caption{M-H plot of (Co,Cu,Mg,Ni,Zn)O system doped with varying iron content}
    \label{fig:MHFe}
\end{figure}

To study the influence of exchange striction on the lattice distortion, 1, 5 and 10 \% iron was added to the ME-TMO to increase the magnetisation and assess the distortion associated with it.  The iron doped ME-TMOs were synthesised by the RCP process. The RCP method was selected for synthesis as it was the most equilibrium process and, furthermore, the peak broadening during X-ray diffraction due to crystallite size effect would be the least, implying that any effect on the XRD peaks would be primarily because of distortion in the lattice. Rietveld refinement of the X-ray diffraction patterns are shown in figure \ref{fig:xrd2}, which confirmed a single phase formation in each case. 
It was observed that the “cubic lattice parameter” increased marginally with increased Fe content and could not be explained only in terms of the presence of a proportional quantity of Fe$^{+2}$ ions in the cation sublattice as the ionic radius of Fe$^{+2}$ (78 pm) is much higher than the ionic radii of the other cations.  This clearly implied that some Fe was present in the cationic sublattice in the +3 oxidation state, whose ionic radius is much lower (56 pm).  M-H plots of Fe doped ME-TMO systems are shown in figure \ref{fig:MHFe}. Magnetic susceptibility at room temperature was calculated from the M-H plots at higher magnetic fields and is listed in table \ref{table:fexrd}.
\begin{table*}[width=.9\textwidth,cols=4,pos=h]
\caption{\label{table:fexrd}Parameters obtained from Rietveld refinement of XRD data and magnetic susceptibility of Fe-doped ME-TMO}
\begin{tabular}{p{2cm}p{2cm}ccccp{2cm}ccp{3.1cm}}
\toprule
 &&\multicolumn{4}{c}{\centering{Monoclinic lattice parameters}}&&&&\\
 \cline{3-6}
 (Co,Ni,Mg,Cu, Zn)$_{1-x}$Fe$_{x}$O &\centering{Cubic lattice parameter (\AA)}&\centering {a (\AA)}&\centering b (\AA)&\centering c (\AA)&\centering {$\beta(^{\circ})$}&Monoclinic cell volume (\AA$^{3}$)&Distortion&Chi$^{2}$&Magnetic susceptibility (10$^{-5}$ (emu g$^{-1}$ Oe$^{-1}$)) \\ \midrule
 X=1&\centering4.232&5.181&2.994&2.991&\centering{125.256}&\centering37.960&0.165&1.62&2.4\\
 X=5&\centering4.234&5.191&2.996&2.998&\centering{125.345}&\centering38.036&0.223&1.26&2.66\\
X=10&\centering4.235&5.194&2.998&2.995&\centering{125.252}&\centering38.096&0.310&1.36&3.08\\
\bottomrule
\end{tabular}
\end{table*}
It is observed that the magnetic susceptibility increased with increase in the iron content. On addition of 1 \% and 5 \% Fe, the system displayed antiferromagnetic behaviour, as a linear M-H trend was observed. However, on addition of 10 \% Fe to the ME-TMO system, a ferromagnetic component was clearly observed at a low magnetic field. Ferromagnetic exchange interaction had resulted from the decreased distance between the iron cations (due to the larger Fe content) in the lattice. 
The lattice distortion (determined by comparing the transformed monoclinic lattice parameters with the ones obtained through Rietveld refinement) increased with increasing iron content.  This showed that exchange striction played a significant role in causing distortion in the structure. However, a comparison of the magnetisation values for the same composition synthesised by different processes revealed that the FSP powders, which had the finest crystallite sizes, also had the lowest magnetisation at room temperature, but the highest distortion. Evidently, when the crystallite sizes of the powders reduced, other factors such as magnetic surface anisotropy started to be of significance. These factors are addressed with further investigations of the magnetic property of the system.
\subsection{\label{sec:5kto300k}Magnetic behaviour from 5 K to 300 K}
\begin{figure}
    \centering
    \includegraphics[width=7.6cm,height=10.5cm]{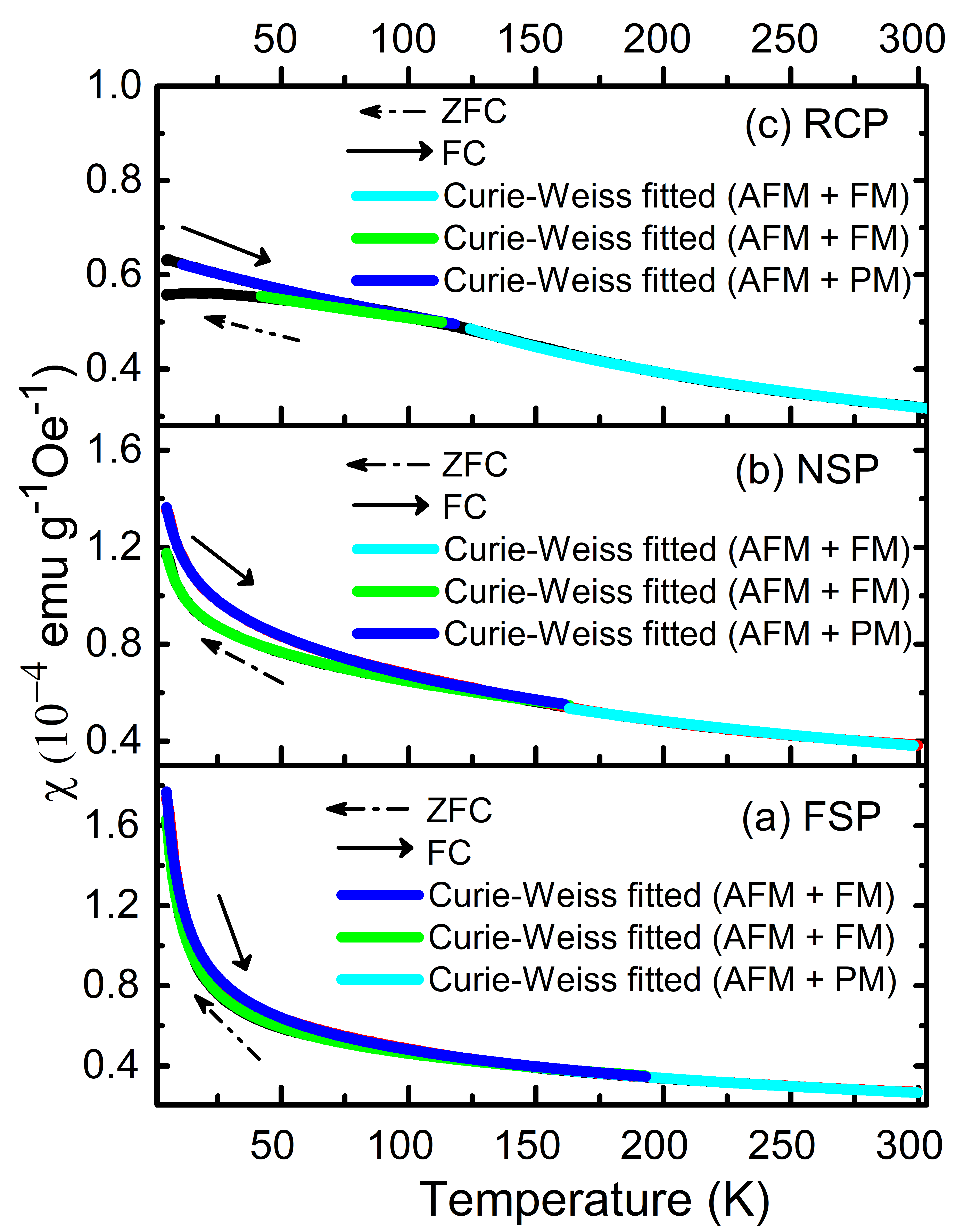}
    \caption{(a), (b), (c) shows the $\chi$-T plot of (Co,Cu,Mg,Ni,Zn)O synthesied by FSP, NSP and RCP respectively and fitted with Curie-Weiss law}
    \label{fig:M-T}
\end{figure}

SQUID-VSM magnetic measurements were carried out to understand the magnetic behaviour from 5 K to room temperature and to find the effect of magnetic anisotropy. $\chi$-T plots of ME-TMO synthesised by FSP, NSP and RCP are shown in figure \ref{fig:M-T}. The Weiss temperature obtained from the magnetisation-temperature measurements by fitting the Curie-Weiss (C-W) law was found to be negative in the full range of measurement from 5 K to 300 K indicating that the ME-TMO was antiferromagnetic (AFM) in nature up to 300 K. The fitted values of the Weiss temperature ($\theta$) are listed in table \ref{table:mag}.  Variation in the Weiss temperature ($\theta$), below and above the bifurcation temperature (T$_{Bif}$) indicated the presence of a magnetic transition at that point. 
Below T$_{Bif}$, the $\chi$-T curve was a sum of two antiferromagnetic susceptibility components (indicated by the subscripts for the Curie constant and the Weiss temperature) as shown in q.~(\ref{eq:CW}), in both zero field cooled (ZFC) and field cooled (FC) measurements of the FSP and NSP powders: 
\begin{equation}
{\chi}=
 \left(
   \frac{2C_{1}}{T-\theta_{1}}
  +\frac{2C_{2}}{T-\theta_{2}}
 \right).
 \label{eq:CW}
\end{equation}
The presence of two components in susceptibility can be explained in terms of a core-shell model \cite{Mandal2009} and is analogous to superparamangetism wherein the outer layer of the nanoparticle has a different magnetic ordering compared to the core due to the reduced co-ordination number on the surface.
\begin{table*}[width=.9\textwidth,cols=9,pos=h]
\caption{\label{tab:table3}Bifurcation temperature, anisotropic constant and parameters obtained from Curie-Weiss fits of $\chi$-T plots measured from 5 K to 300 K and P$_{asy}$ calculated from EPR spectroscopy}
\begin{tabular}{p{1.2cm}p{1.2cm}p{1.2cm}p{1.2cm}p{1.2cm}p{1.2cm}p{2cm}p{1.2cm}p{1.2cm}}
\toprule
 &\multicolumn{2}{c}{ZFC\textsuperscript{a}}&\multicolumn{2}{c}{FC\textsuperscript{a}}&&\\
 \cline{2-3}\cline{4-5}
 Process&$\theta_{1}$(K) 
&$\theta_{2}$(K)&$\theta_{1}$(K)
&$\theta_{2}$(K)&T$_{Bif}$ (K)& K (erg/cm$^{3}$)&$\theta$(K)\textsuperscript{b}&P$_{asy}$
\\ \midrule
 FSP&-1.97&-291.37&-2.16&-198.39&183&357452.2&-167.46&2.33\\
 NSP&-1.18&-263.64&-2.56&-196.03&165&69615.2&-173.17&2.02\\
 RCP&\centering-&\centering-&-11.80&-415.16&144&1787.2&-219.50&1.93\\
\bottomrule
\end{tabular}
\begin{flushleft}
    \textsuperscript{a}{Weiss temperature from 5 K to bifurcation temperature (T$_{Bif}$)}\\
    \textsuperscript{b}{Weiss temperature from bifurcation temperature (T$_{Bif}$) to 300 K}
  \label{table:mag} 
  \end{flushleft}
\end{table*}

\begin{figure}
    \centering
    \includegraphics[width=8.6cm,height=4cm]{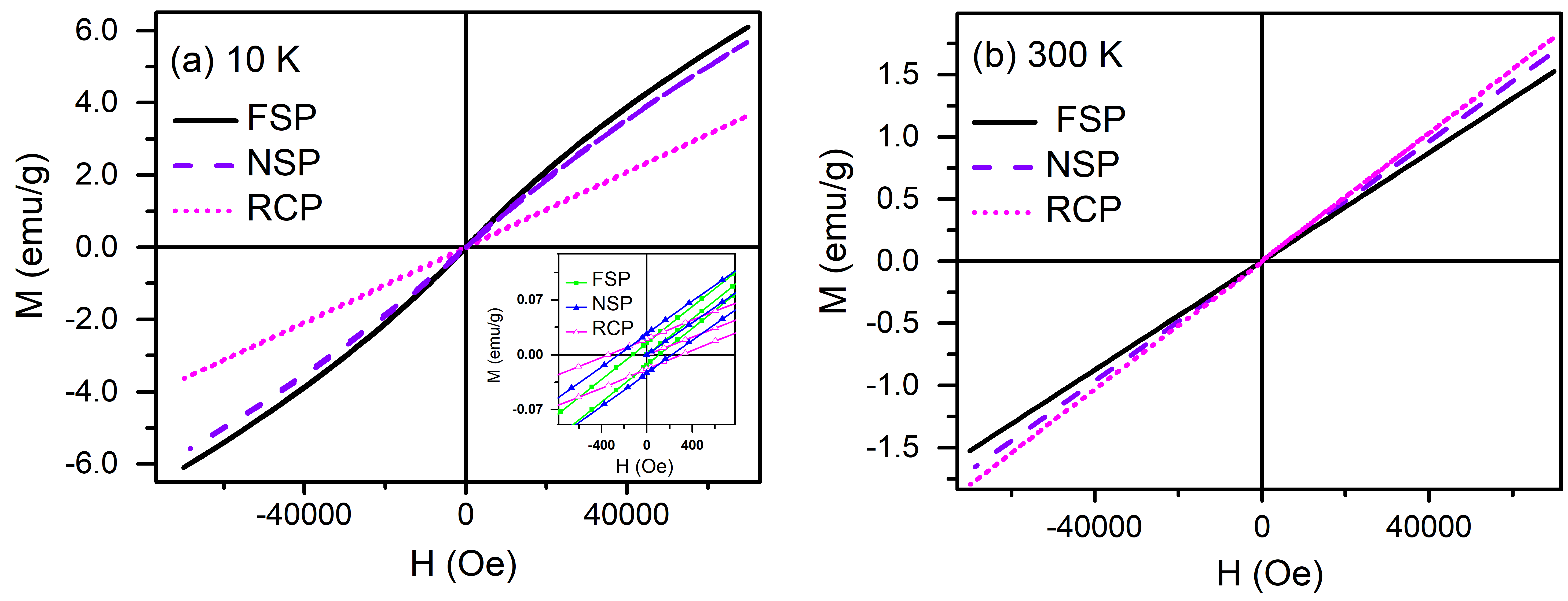}
    \caption{(a) shows M-H plot of (Co,Cu,Mg,Ni,Zn)O powders at 10 K with inset showing exchange bias (b) shows M-H plot of all the process at 300 K}
    \label{fig:MH300}
\end{figure}

One component of the magnetic susceptibility ((2C$_{1}$)/(T-$\theta_{1}$)) had a high negative Weiss temperature ($\theta_1$) indicating a strong AFM ordering as $\theta$ is directly proportional to the exchange interaction (J), and originated from the core of the crystallite. On the other hand, the other component ((2C$_{2}$)/(T-$\theta_{2}$)) was weakly antiferromagnetic, which was due to the presence of coupled antiferromagnetic and ferromagnetic (FM) interactions in the shell of the crystallite. AFM ordering has a negative J while FM ordering has a positive J, and hence, the presence of FM exchange led to a decrease in the overall J value which, in turn, resulted in a reduced value of the Weiss temperature.  Presence of ferromagnetism can be also confirmed from the M-H plot of the FSP powders at 10 K (shown in figure \ref{fig:MH300})  as an ‘S’ type curve with a change of slope at the spin flop field around 10000 Oe. Since the M-H loop did not saturate at high fields, the existence of both antiferromagnetic and ferromagnetic ordering in the system could be deduced, which also supported the core-shell model.\\ 

\begin{figure}
    \centering
    \includegraphics[width=6.6cm,height=5.5cm]{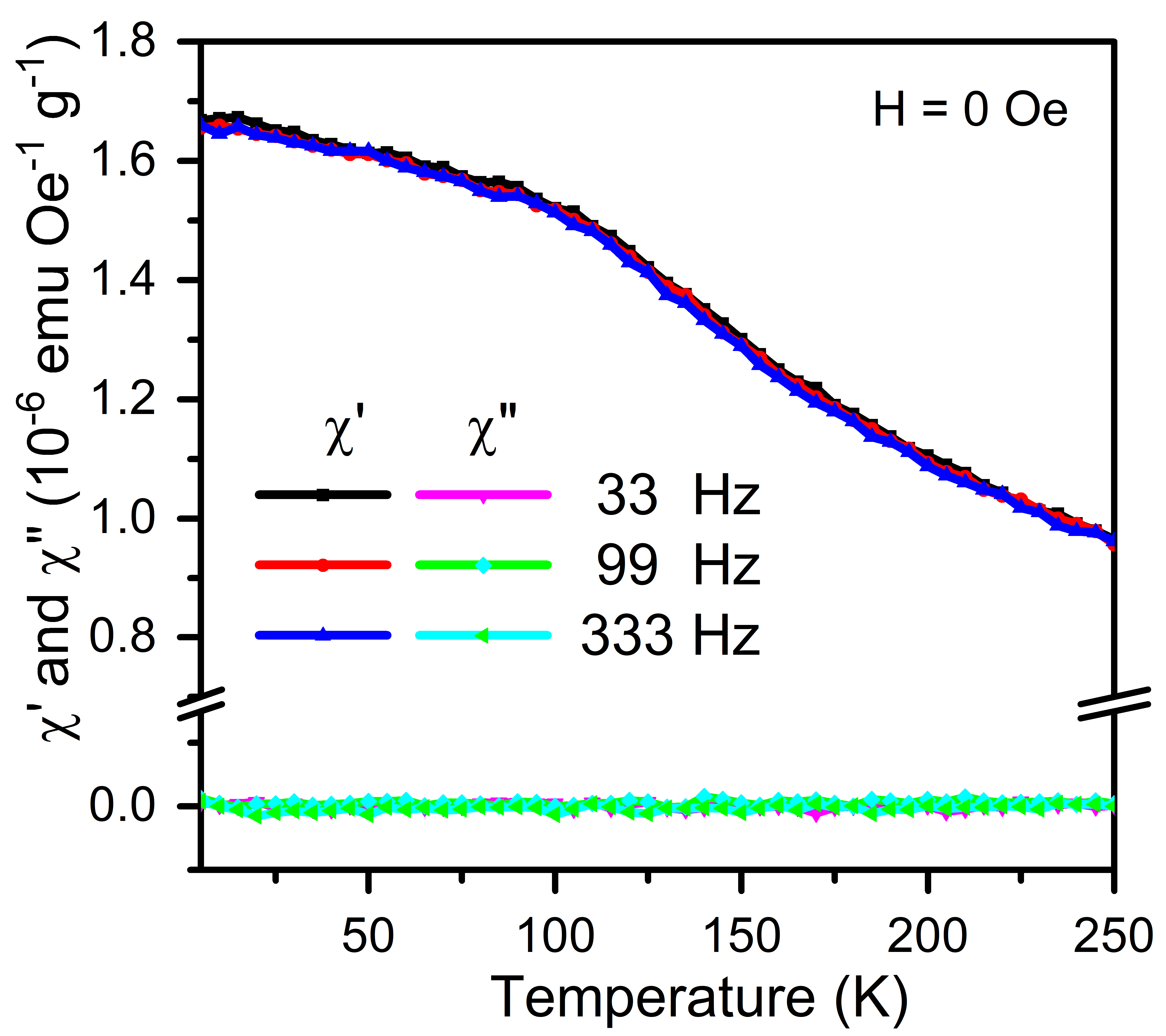}
    \caption{ AC magnetic susceptibility plot of (Co,Cu,Mg,Ni,Zn)O synthesised through RCP process}
    \label{fig:AC_RCP}
\end{figure}

The powders synthesised by RCP also showed a similar trend under FC conditions, but ZFC conditions showed a non Curie-Weiss behaviour.  To assess the presence of spin glass behaviour, AC magnetic susceptibility measurements were carried out which showed absence of frequency dependence in the susceptibility measurement. This implied that that spin glass behaviour was not present in the lattice. AC magnetic susceptibility plot of the RCP powder is shown in figure \ref{fig:AC_RCP}. On the other hand during FC conditions, in the presence of a magnetic field, ordering of the spins became possible and hence, a weak antiferromagnetic component was present in the shell along with strong AFM ordering in the core. However, the ferromagnetic exchange interaction in the RCP powders was less compared to the FSP powders which can be seen from the Weiss temperature value below T$_{Bif}$.\\ 

Core-shell magnetic behaviour in both FSP and RCP powders was also confirmed from the presence of exchange bias, \cite{Makhlouf1997} exhibited by the shift of the M-H loop towards the negative axis as seen in figure  \ref{fig:MH300}. The relative volumes of the core and shell for the crystallites synthesised by FSP, NSP and RCP were calculated from the deconvoluted susceptibilities found from C-W law fits and correlated with the crystallite sizes found from x-ray diffraction. It was estimated that the FSP powders had 59.2 \% of the volume as the shell, the NSP powders had 23.5 \% while only 1 \% of the volume was the shell in case of the RCP powders. The relatively larger proportion of the shell volume in case of FSP resulted in a higher magnetisation value below the T$_{Bif}$. Above T$_{Bif}$, however, for all the three powders, an antiferromagnetic behaviour was obtained by describing the susceptibility by a simple Curie-Weiss law. Among the three powders, the magnetisation of the FSP powders was found to be the lowest beyond the blocking temperature. This observation was attributed to the smaller size of the core, while the shell transformed from ferromagnetic to paramagnetic behaviour above the blocking temperature, as can be observed from the M-H plots at room temperature.
The coercivity was calculated using the Eq.~(\ref{eq:Hc}),
\begin{equation}
\left\{
H_{c}=\frac{H^{+}-H^{-}}{2}
\right\}.
\label{eq:Hc}
\end{equation}
where H$^{+}$ is the switching field on the right side and H$^{-}$ on the left side of the hysteresis loop.\cite{Sun2005} H$_{c}$ of the FSP powders was 122.6 Oe at 10 K, 241.5 Oe for NSP while that for RCP was 319.5 Oe.  Increased coercivity in the NSP and RCP powders was attributed to the larger antiferromagnetic core size. Presence of weak ferromagnetic behaviour in the shell could be due to the canting of the spins, where the spins do not cancel each other completely due to change in the angle of the spins\cite{Peck2011} or due to the presence of weak ferromagnetism (FM) from uncompensated spins at the surface.\cite{Meneses2010}\\

\begin{figure}
    \centering
    \includegraphics[width=5.6cm,height=7cm]{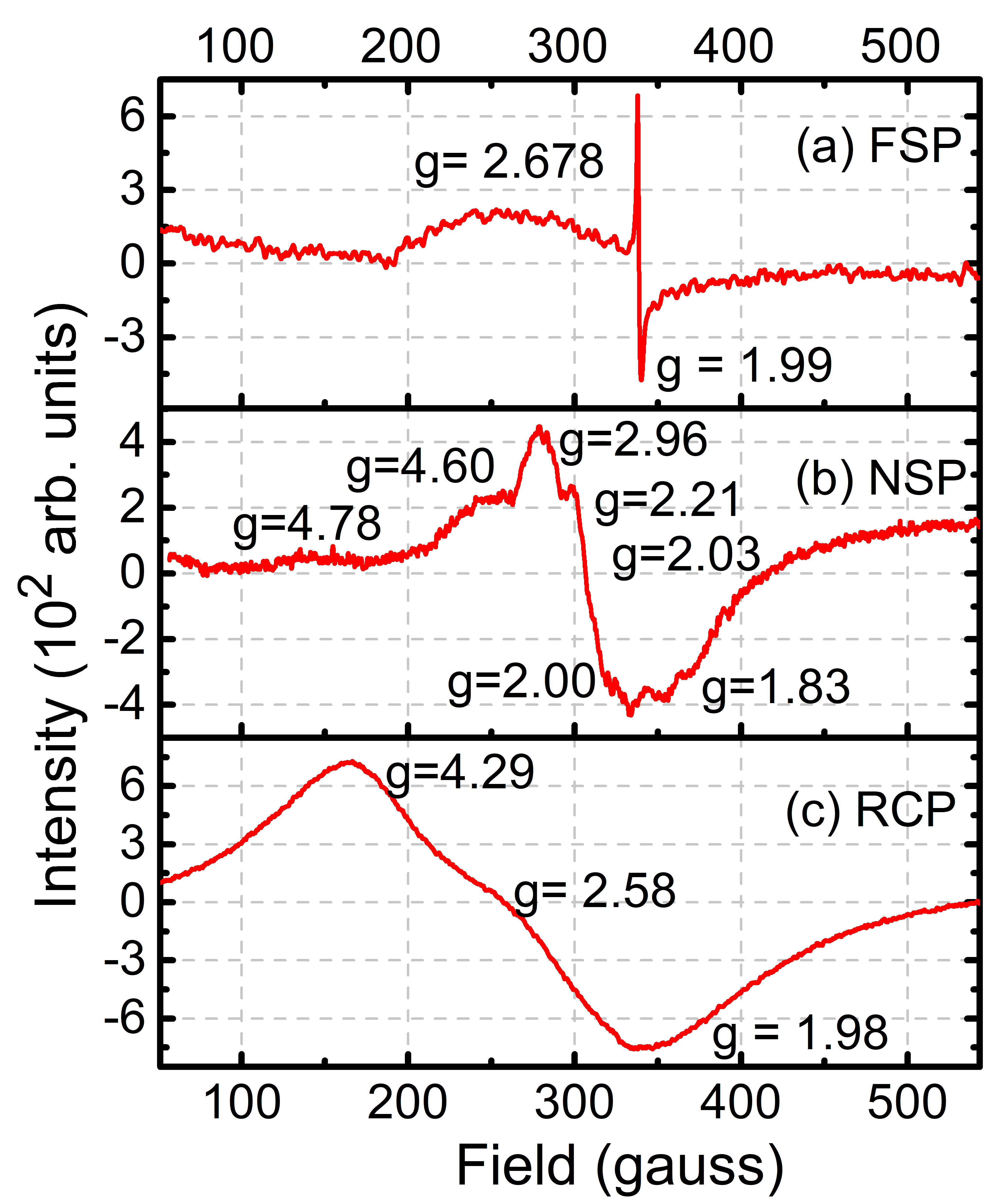}
    \caption{Room temperature EPR spectra of (Co,Cu,Mg,Ni,Zn)O synthesised through (a) FSP, (b) NSP and (c) RCP powders}
    \label{fig:EPR}
\end{figure}

\begin{figure}
    \includegraphics[width=7.6cm,height=3.55cm]{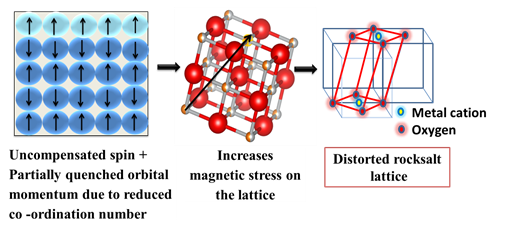}
    \caption{Schematic of distortion from cubic to monoclinic lattice}
    \label{fig:Schem}
\end{figure}

Presence of ferromagnetism was observed from electron spin resonance (ESR)/electron paramagnetic resonance (EPR) spectroscopy. Figure \ref{fig:EPR} shows the ESR/EPR spectra for the powders obtained from the three processes. 
Asymmetry of the peak indicated the presence of magnetic anisotropy in the lattice while broadening showed the presence of ferromagnetic interaction at room temperature. The hyperfine structure observed in the NSP powders indicated higher distortion compared to RCP and the consequent lifting of orbital degeneracy led to anisotropic g factors which were calculated from the Eq.~(\ref{eq:EPR}),\cite{Boukhari2019}
\begin{equation}
\left\{
h\mu = g\beta H_r%
\right\}.
\label{eq:EPR}
\end{equation}
where h is Planck’s constant, $\mu$ is the microwave frequency, $\beta$ is the Bohr magneton radius and H$_r$ is the resonance magnetic field. In the case of FSP powders, the sharp spike in EPR resonance indicates the presence of an increased amount of carrier concentrations, where the electrons behave as a Fermi gas, leading to narrow resonance.\\
Asymmetric features at a lower field shows the presence of a ferromagnetic component.\cite{He2015} Presence of spin canting was analysed from the EPR measurements. As the general Hamiltonian which describes the symmetric exchange cannot account for canted spins, the antisymmetric exchange, also known as the Dzyaloshinskii-Moriya interaction (D-M) was considered.  $\mathcal{H}_{c}$ describes the Hamiltonian of the system where weak ferromagnetic exchange could occur due to the canting of the spins and is given in Eq.~(\ref{eq:Ham}),
\begin{equation}
\left\{
\mathcal{H}_{c} = \mathbf D_{ij} \cdot \mathbf [\mathbf S_{i} \times \mathbf S_{j}]%
\right\}.
\label{eq:Ham}
\end{equation}
where D$_{ij}$ is the antisymmetric exchange vector, and S$_{i}$ and S$_{j}$ are the two neighbouring magnetic spins of the interacting electrons.\cite{Carlin1935} Spin canting occurs due to increased magnetic anisotropy arising from the presence of cations with different directions of easy axis or due to the  surface anisotropy induced by the synthesis process. Spin canting can be studied using D$_{ij}$, as it depends on the immediate symmetry surrounding two atoms in the lattice. When there is spin canting in the lattice, anisotropy increases, which can be indicated by the variation in D$_{ij}$, and is proportional to (g-2)/2. However, resonance in case of powders from all three processes did not occur at the same field strength and hence, it was not possible to estimate the spin canting from D$_{ij}$. Nevertheless, magnetic anisotropy can be estimated from P$_{asy}$=(1-(h$_{u}$/h$_{l}$)), where h$_{u}$ is the height of the upper peak and h$_{l}$ is height of the lower peak.\cite{Das2017} Table  \ref{table:eprraman} gives the values for each of the synthesis processes. It can be seen that the FSP powders had the highest magnetic anisotropy which was in agreement with SQUID measurements.\\
Increased magnetic anisotropy in the FSP powders can be explained as follows. Typically in transition metal compounds, the orbital angular momentum is quenched due to the crystal field effect. On the other hand, in nanoparticles, the reduced co-ordination number (because of the larger specific surface area) leads to only partial quenching of the orbital angular momentum, which, in turn, increases the surface magnetic anisotropy in the material.\cite{Golosovsky2005} To reduce the surface magnetic anisotropy, spin canting takes place. Since the FSP powders had the highest surface anisotropy due to the increased specific surface area, it exhibited the highest bifurcation temperature. Due to the increased magnetic anisotropy at the surface, core-shell type of magnetic behaviour was observed. Increased surface magnetic anisotropy also led to increased magnetic stress in the material or domain formation in the surface.\cite{Gomonay2007} The repulsion between the magnetic domains would have caused magnetostriction in the nanocrystalline ME-TMO and would result in greater distortion in the crystal structure in case of the powders with the finer crystallite sizes, as confirmed from analysis of X-ray diffraction data. Figure \ref{fig:Schem} shows a schematic of the distortion mechanism.
\begin{table}[width=.5\textwidth,cols=4,pos=h]
\caption{\label{tab:table1}Phonon life, magnon position from Raman spectra and J interaction calculated from VSM measurements }
\begin{tabular}{p{5.4cm}p{0.6cm}p{0.6cm}p{0.6cm}}
\toprule
Parameter&FSP&NSP&RCP\\
\hline
Phonon life (10$^{-14}$ s) & 5.03 & 5.12 & 5.91\\
Magnon peak position (cm$^{-1}$) & 1635 & 1604 & 1589\\
J interaction from Neel temperature (K) & 197.64 & 196.80 & 173.18\\
Magnetic transition temperature (K) & 703 & 700 & 616\\
\bottomrule
\label{table:eprraman}
\end{tabular}
\end{table}\\
Neel-Brown has developed the relation between magnetic anisotropy and blocking temperature (T$_b$) as given by Eq.~(\ref{eq:aniso})
\begin{equation}
\left\{
 K = \frac{25k_{b}T_{b}}{V}%
\right\}.
\label{eq:aniso}
\end{equation}
where K is the magnetic anisotropy, V is volume of the crystallite calculated from the crystallite size obtained from XRD and k$_{b}$ is the Boltzmann constant.\cite{Meneses2010}\\

Bifurcation temperature can be considered to be the highest blocking temperature when there is a distribution in the crystallite sizes.\cite{Montes2014} The bifurcation temperatures obtained in the present study are given in table \ref{table:mag} and was found to be the highest for the FSP powders and least for the RCP powders. Although, in general, the blocking temperature decreases with a decrease in the crystallite size due to the reduced co-ordination number,\cite{Montes2014} in the present case a reverse trend was observed. Such behaviour of increase in the bifurcation temperature has been exhibited by NiO nanoparticles. \cite{Thota2007} These phenomena can be explained in terms of increased surface magnetic anisotropy leading to an increased bifurcation temperature, as given by the Neel-Brown equation (Eq.~(\ref{eq:aniso})) wherein the blocking temperature is directly proportional to the anisotropy. Magnetic anisotropy (K) calculated from the Eq.~(\ref{eq:aniso}) is given in table \ref{table:mag}, confirming that FSP powders had the highest magnetic anisotropy in agreement with the EPR result.\\

SQUID measurements confirmed the system to be antiferromagnetic at room temperature. However, the Neel temperature of the ME-TMO has been reported in the literature as 113 K,\cite{Jimenez-segura2019, Zhang2019, Mao2019} indicated by a cusp like feature in the $\chi$-T plot, while at the same time short range antiferromagnetic ordering and magnetic excitation continued up to room temperature.\cite{Zhang2019} The RCP powders showed a transition with a cusp like a feature that was not as prominent as reported in literature. On the other hand, the FSP and NSP powders did not show any such transition. This could be attributed to the crystallite size effect and increased defect states in the FSP and NSP powders when compared to that synthesised by RCP. Defect states, quantified from Raman spectroscopy, showed that FSP powders had the highest defect state concentration while the RCP powders had the least.  Exchange interaction of the cations with metal vacancies leads to the stabilisation of antiferromagnetism above the reported Neel temperature. 

\subsection{\label{sec:HTmag}High temperature magnetic behaviour}

\begin{figure}
\centering
    \includegraphics[width=8.6cm,height=7.6cm]{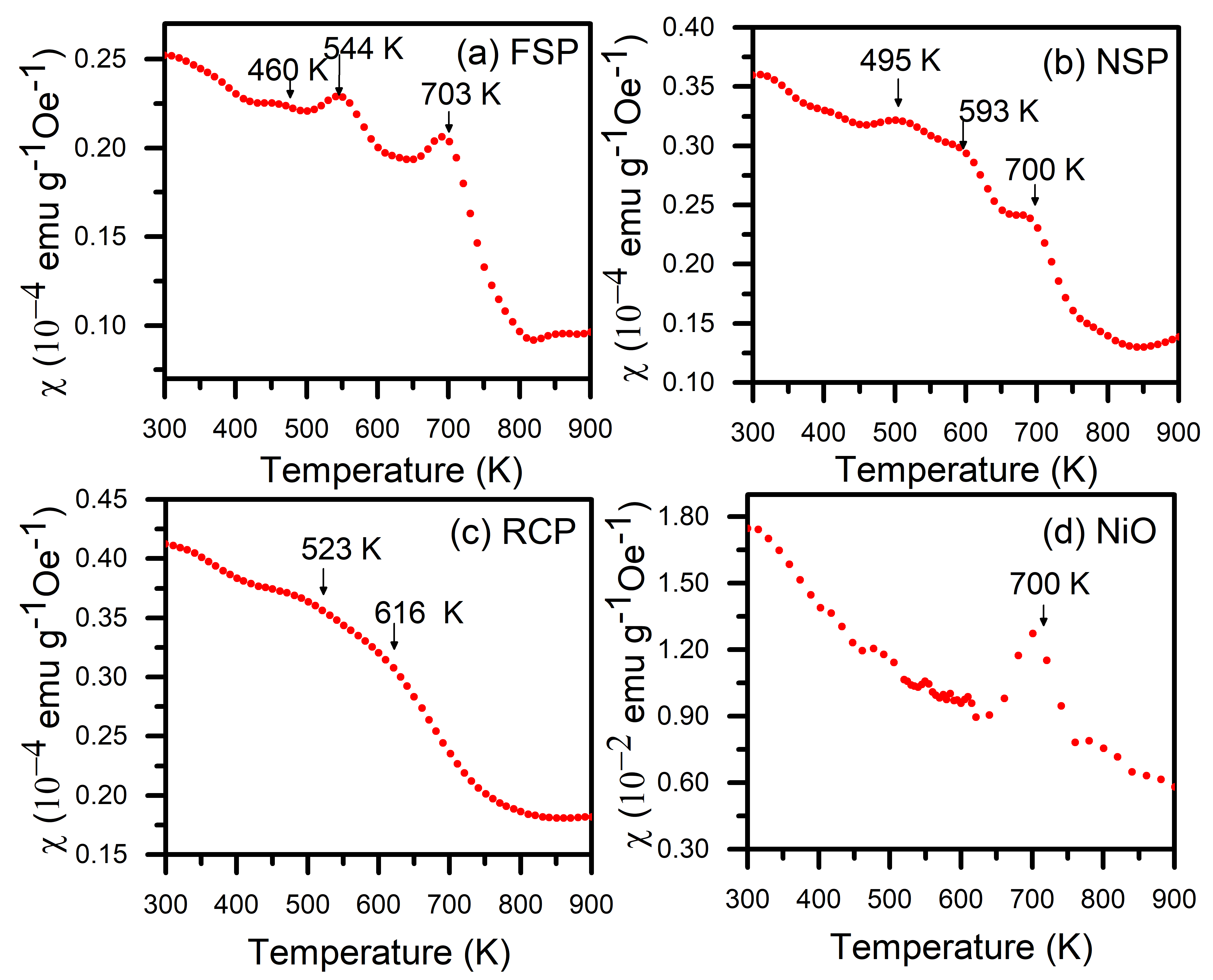}
    \caption{(a)-(c) show high temperature $\chi$-T plot of (Co,Cu,Mg,Ni,Zn)O synthesised by FSP, NSP and RCP respectively, and (d) shows $\chi$-T plot of NiO synthesised by NSP}
    \label{fig:highm-t}
\end{figure}

Magnetic transitions above room temperature was characterised by VSM measurements from 300 K to 800 K. $\chi$-T plots, shown in figure  \ref{fig:highm-t}, reveal sharp transitions at 703 K, 700 K and 612 K for the FSP, NSP and RCP powders respectively. Curie-Weiss law fits indicated paramagnetic behaviour in all three cases confirming the transitions to be the respective Neel temperatures for the powders. These were even higher than that of bulk NiO (523 K) which has the highest Neel temperature among the individual oxides present in the system. Increase in the Neel temperature of a material could arise due to the presence of increased defect states.  For confirmation, nanocrystalline NiO was synthesised by NSP and $\chi$-T characterisation carried out (figure \ref{fig:highm-t}(d)). It can be seen that the Neel temperature of the synthesised NiO was 700 K (much higher than the previously reported 523 K) and arose from the increased defect states created during synthesis.  An increase in the Neel temperature has been reported in ball-milled NiO and Ni-doped in TiO$_{2}$.\cite{Ravikumar2015}  In the case of MnO, increase in the Neel temperature was attributed to the interactions of the spins with defect states and the ferromagnetic shell with the AFM core.\cite{Golosovsky2005} The other transitions observed below the Neel temperature in figure \ref{fig:highm-t} can be attributed to the transition temperatures of Ni superexchange interactions with the other cations.

The J$_{2}$ (next nearest neighbour) interaction was calculated from the molecular field approximation relation,\cite{Joshi2018,Bracconi1983} as given in the Eq.~(\ref{eq:neel}),
\begin{equation}
\left\{
T_{N} = 2J_{2}[(S_{Cu})(S_{Cu}+1)(S_{Ni})(S_{Ni}+1)(S_{Co})(S_{Co}+1)]^{1/3}
\right\}.
\label{eq:neel}
\end{equation}
where S is the spin of the electrons. The calculated values are listed in table \ref{table:eprraman} where it can be seen that the exchange interaction was higher for the FSP and NSP powders when compared to RCP. Since the magnetic anisotropy was also high in the FSP powders, the structure had the highest distortion at room temperature. 
The J$_2$  interaction is directly proportional to the magnon peak position in the Raman spectra\cite{Funkenbusch1981} and the trend in the values of the magnon peak position with respect to the process is similar to the J$_2$  interaction values, which confirmed that FSP had the highest J$_2$  interaction and Neel temperature.

\subsection{\label{sec:JahnTeller}Effect of Jahn-Teller distortion and electronegativity }
The distortion in the system could have also been caused by other factors like the Jahn-Teller effect. Among the principal constituents, only copper oxide shows Jahn-Teller distortion in its structure due to the crystalline field effect. Bérardan \textit{et al.}\cite{Berardan2017} showed that decreasing the copper content in ME-TMO decreases the distortion in the structure. Rost \textit{et al.}\cite{Rost2017} also showed a variation in the bond length near the Cu cation. However, when copper was eliminated from the parent ME-TMO and a (Co,Mg,Ni,Zn)O system was synthesised by FSP, a distortion in the structure was, nonetheless, evident from the asymmetry in the XRD peaks, as seen in figure\ref{fig:XRD}(d). The lattice parameter deviation from the ideal cubic structure is shown in table \ref{table:xrd}. Hence, it was surmised that Jahn-Teller distortion introduced due to copper was not the only source for variation from the cubic structure. 
Another fact to be considered is that each unary oxide has a different bond length. The principal cations have different electronegativities and so, the ionicities of the bonds would vary. The bond length would be shorter for a covalent bond when compared to an ionic bond and this variation would cause the lattice to have a distorted structure\cite{Goodenough1955} as well as a volume striction in the lattice. 

\section{Conclusions}

A phase-pure transition metal oxide based multicomponent equimolar (Co,Cu,Mg,Ni,Zn)O system was shown to have a distorted cubic or monoclinic structure rather than the ideal rocksalt structure, regardless of the synthesis process. Magnetic characterisation indicated that the distortion was predominantly due to the presence of antiferromagnetism, originating from the principal constituents such as NiO and CoO. Addition of Fe to the lattice increased the distortion due to exchange striction. It was shown that a smaller crystallite size with increased number of defect states introduced more uncompensated spin states, spin canting and partial quenching of orbital momentum leading to varied magnetic behaviour at the surface and the interior of the crystallite resulting in a “core-shell” magnetic behaviour. This magnetic anisotropy resulted in the distortion of the cubic lattice. Interaction of spin with the defect states also resulted in a higher Neel temperature when compared to bulk NiO which has the highest Neel temperature among the constituent unary oxides. Presence of distortion in a system synthesised without Cu implied that the Jahn-Teller effect was not the only cause. Additionally, the varying ionicity and length of the metal cation – oxygen bonds would also contribute to the observed volume striction in the lattice. 

\section*{Acknowledgments}
We acknowledge the DST, Government of India (GoI) for funding the NFMTC where this work was carried out and SAIF, IIT Madras for the EPR, VSM facilities, The Nanotechnology lab, MME department for the XRD measurements and Physics department, IIT Madras for SQUID measurement. Abhishek Sarkar, KIT, Germany is gratefully acknowledged for confocal micro-Raman spectroscopy measurements. The support from a DST, SERB project\\
SERB/F/4352/2017-2018 is gratefully acknowledged.

\appendix
\section{Appendix}
Supplementary information provides, d-spacing and hkl planes of each reflection of XRD pattern in table S1-S2. TEM micrograph of (Co,Cu,Mg,Ni,Zn)O synthesied through FSP, NSP and RCP is shown in figure S1. EDS elemental mapping of (Co,Cu,Mg,Ni,Zn)O  synthesised through FSP, NSP and RCP and Fe doped (Co,Cu,Mg,Ni,Zn)O  is shown in figure S2-S7 .

\printcredits



\nocite{*}

\begin{thebibliography}{53}

\bibitem{Rost2015} C.M. Rost, E. Sachet, T. Borman, A. Moballegh, E.C. Dickey, D. Hou, J.L. Jones, S. Curtarolo, J.P. Maria, 2015. Entropy-stabilized oxides, Nat. Commun. 6, 8485. doi:https://doi.org/10.1038/ncomms9485
\bibitem{Berardan2016} D. Bérardan, S. Franger, D. Dragoe, A.K. Meena, N. Dragoe, Colossal dielectric constant in high entropy oxides, Phys. Status Solidi - Rapid Res. Lett. 10 (2016) 328–333. doi:https://doi.org/10.1002/pssr.201600043
\bibitem{Sarkar2018} A. Sarkar, L. Velasco, D. Wang, Q. Wang, G. Talasila, L. De Biasi, C. Kübel, T. Brezesinski, S.S. Bhattacharya, H. Hahn, B. Breitung, 2018. High entropy oxides for reversible energy storage, Nat. Commun. 9, 3400. doi:https://doi.org/10.1038/s41467-018-05774-5
\bibitem{Berardan2017} D. Bérardan, A.K. Meena, S. Franger, C. Herrero, N. Dragoe, Controlled Jahn-Teller distortion in (MgCoNiCuZn)O-based high entropy oxides, J. Alloys Compd. 704 (2017) 693–700. doi:https://doi.org/10.1016/j.jallcom.2017.02.070
\bibitem{Smart1951} J.S. Smart, S. Greenwald, Crystal structure in antiferromagnetic compound at the Curie temperature, Phys. Rev. 82 (1951) 113–114. doi:https://doi.org/10.1103/PhysRev.82.113
\bibitem{Derakhshan2008} S. Derakhshan, J.E. Greedan, T. Katsumata, L.M.D. Cranswick, Long-range antiferromagnetic ordering in the novel magnetically frustrated rock salt oxide system: Li$_3$Mg$_2$RuO$_6$, Chem. Mater. 20 (2008) 5714–5720. doi:https://doi.org/10.1021/cm801161r
\bibitem{Dalverny2010} A.L. Dalverny, J. Filhol, F. Lemoigno, M.L. Doublet, Interplay between magnetic and orbital ordering in the strongly correlated cobalt oxide: A DFT+U Study, J. Phys. Chem. C. 114 (2010) 21750–21756. doi:https://doi.org/10.1021/jp108599m
\bibitem{Roth.W.L1958} Roth. W. L, Magnetic Structures of MnO, FeO, CoO, and NiO, Phys. Rev. 110 (1958) 1333–1341. doi:https://doi.org/10.1103/PhysRev.110.1333
\bibitem{Lee2016} S. Lee, Y. Ishikawa, P. Miao, S. Torii, T. Ishigaki, T. Kamiyama, 2016. Magnetoelastic coupling forbidden by time-reversal symmetry: Spin-direction-dependent magnetoelastic coupling on MnO, CoO, and NiO, Phys. Reivew B. 93,064429. doi:https://doi.org/10.1103/PhysRevB.93.064429
\bibitem{Li1955} Y.-Y. Li, Magnetic moment arrangement and magnetocrystalline deformation in antiferromagnetic compounds, Phys. Rev. 100 (1955) 627–631. doi:https://doi.org/10.1103/PhysRev.100.627
\bibitem{Rooksby1948} H.P. Rooksby, 1948. A note on the structure of nickel oxide at subnormal and elevated temperatures., Acta Crystallogr. 1, 226. doi:https://doi.org/10.1107/S0365110X48000612
\bibitem{Kanamori1957} J. Kanamori, Theory of the Magnetic properties of ferrous and cobaltous oxides, II, Prog. Theor. Phys. 17 (1957) 197–222. doi:https://doi.org/10.1143/PTP.17.197
\bibitem{Greenwald1950} S. Greenwald, J.S. Smart, Deformation in th crystal structures of anti-ferromagnetic Compounds, Nature. 166 (1950) 523–524. doi:https://doi.org/10.1038/166523a0
\bibitem{Bean1962} C.. Bean, D.. Rodbell, Magnetic disorder as a first-order phase transformation, Phys. Rev. 126 (1962) 104–115. doi:https://doi.org/10.1103/PhysRev.126.104
\bibitem{Bartel1970} L.C. Bartel, Antiferromagnetism in MnO Calculation of Near-Neighbor Spin Correlation Functions for T<TN, Phys. Rev. B. 1 (1970) 1254–1260. doi:https://doi.org/10.1103/PhysRevB.1.1254
\bibitem{Rinaldi-montes2016} N. Rinaldi-montes, P. Gorria, D. Martínez-blanco, A.B. Fuertes, I. Puente-orench, J.A. Blanco, 2016. Size effects on the Néel temperature of antiferromagnetic NiO nanoparticles, AIP Adv. 6, 056104. doi:https://doi.org/10.1063/1.4943062
\bibitem{Kremenovic2012} A. Kremenovic, B. Jancar, M. Ristic, M. Vuc, J. Rogan, A. Pacevski, B. Antic, Exchange-Bias and grain-surface relaxations in nanostructured NiO/Ni induced by a particle size reduction, J. Phys. Chem. C. 116 (2012) 4356–4364. doi:https://doi.org/10.1021/jp206658v
\bibitem{Golosovsky2001} I. V Golosovsky, I. Mirebeau, G. André, D.A. Kurdyukov, Y.A. Kumzerov, S.B. Vakhrushev, Magnetic ordering and phase transition in MnO embedded in a porous glass, Phys. Review Lett. 86 (2001) 5783–5786. doi:https://doi.org/10.1103/PhysRevLett.86.5783
\bibitem{Sarkar2017} A. Sarkar, R. Djenadic, N.J. Usharani, K.P. Sanghvi, V.S.K. Chakravadhanula, A.S. Gandhi, H. Hahn, S.S. Bhattacharya, Nanocrystalline multicomponent entropy stabilised transition metal oxides, J. Eur. Ceram. Soc. 37 (2017) 747–754. doi:https://doi.org/10.1016/j.jeurceramsoc.2016.09.018
\bibitem{Duan2012} W.J. Duan, S.H. Lu, Z.L. Wu, Y.S. Wang, Size effects on properties of NiO nanoparticles grown in alkalisalts, J. Phys. Chem. C. 116 (2012) 26043–26051. doi:https://doi.org/10.1021/jp308073c
\bibitem{Usharani2019}N.J. Usharani, S.S. Bhattacharya, Effect of defect states in the optical and magnetic properties of nanocrystalline NiO synthesised in a single step by an aerosol process, Ceram. Int. (2019). doi:https://doi.org/10.1016/j.ceramint.2019.11.014 
\bibitem{Mironova-ulmane2008} N. Mironova-ulmane, U. Ulmanis, A. Kuzmin, I. Sildos, M. Pärs, C.M. Guidi, M. Piccinini, A. Marcelli, Magnetic Ordering in Co$_c$Mg$_{1–c}$O Solid Solutions, Proc. XIII Feofilov Symp. “Spectroscopy Cryst. Doped by Rare-Earth Transit. Met. Ions.” 50 (2008) 1723–1726. doi:https://doi.org/10.1134/S1063783408090266
\bibitem{Mironova-Ulmane2007} N. Mironova-Ulmane, A. Kuzmin, I. Steins, J. Grabis, I. Sildos, M. Pars, 2007. Raman scattering in nanosized nickel oxide NiO, J. Phys. Conf. Ser. 93, 012039. doi:https://doi.org/10.1088/1742-6596/93/1/012039
\bibitem{Chou1976} H. Chou, H.Y. Fan, Light scattering by magnons in CoO, MnO, and MnS, Phys. Rev. B. 13 (1976) 3924–3938. doi:https://doi.org/10.1103/PhysRevB.13.3924
\bibitem{Bergman1999} L. Bergman, D. Alexson, P.L. Murphy, R.J. Nemanich, M. Dutta, M. Stroscio, C. Balkas, H. Shin, R.F. Davis, Raman analysis of phonon lifetimes in AlN and GaN of wurtzite structure, Phys. Rev. B. 59 (1999) 977–982. doi:https://doi.org/10.1103/PhysRevB.59.12977
\bibitem{Chanda2017} A. Chanda, S. Gupta, M. Vasundhara, S.R. Joshi, G.R. Mutta, J. Singh, 2017. Study of structural, optical and magnetic properties of cobalt doped ZnO nanorods, RSC Adv. 7, 50527. doi:10.1039/C7RA08458G
\bibitem{Ali2019} N. Ali, B. Singh, Z.A. Khan, A.R. Vijaya, K. Tarafder, S. Ghosh, 2019. Origin of ferromagnetism in Cu-doped ZnO, Sci. Rep. 9, 2461. doi:https://doi.org/10.1038/s41598-019-39660-x
\bibitem{Xu2010} Q. Xu, J. Yang, Magnetism of Zn$_{0.98}$Co$_{0.0}$2O nanopowders, Phys. B. 405 (2010) 1216–1220. doi:10.1016/j.physb.2009.11.045
\bibitem{Theyvaraju2015} D. Theyvaraju, S. Muthukumaran, Preparation, structural, photoluminescence and magnetic studies of Cu doped ZnO nanoparticles co-doped with Ni by sol – gel method, Phys. E. 74 (2015) 93–100. doi:https://doi.org/10.1016/j.physe.2015.06.012
\bibitem{Coey2005} J.M.D. Coey, M. Venkatesan, C.B. Fitzgerald, Donor impurity band exchange in dilute ferromagnetic oxides, Nat. Mater. 4 (2005) 173–179. doi:10.1038/nmat1310
\bibitem{Bartel1971} L.C. Bartel, B. Morosin, Exchange striction in NiO, Phys. Reivew B. 3 (1971) 1039–1042. doi:https://doi.org/10.1103/PhysRevB.3.1039
\bibitem{Jimenez-segura2019} Jimenez-Segura, M.P., Takayama, T., Bérardan, D., Hoser, A.,Reehuis, M., Takagi, H., Dragoe, N., 2019. Long-range magnetic ordering in rocksalt-type high-entropy oxides. Applied Physics Letters 114, 122401. doi:10.1063/1.5091787,
\bibitem{Zhang2019} J. Zhang, J. Yan, S. Calder, Q. Zheng, M.A. McGuire, D.L. Abernathy, Y. Ren, S.H. Lapidus, K. Page, H. Zheng, J.W. Freeland, J.D. Budai, R.P. Hermann, long-range antiferromagnetic order in a rocksalt high entropy oxide, Chem. Mater. 31 (2019) 3705–3711. doi:https://doi.org/10.1021/acs.chemmater.9b00624
\bibitem{Mandal2009} S. Mandal, S. Banerjee, K.S.R. Menon, 2009. Core-shell model of the vacancy concentration and magnetic behavior for antiferromagnetic nanoparticle, Phys. Reivew B. 80, 214420. doi:https://doi.org/10.1103/PhysRevB.80.214420
\bibitem{Makhlouf1997} S.A. Makhlouf, F.T. Parker, F.E. Spada, A.E. Berkowitz, Magnetic anomalies in NiO nanoparticles, J. Appl. Phys. 81 (1997) 5561. doi:https://doi.org/10.1063/1.364661
\bibitem{Sun2005} L. Sun, P.C. Searson, C.L. Chien, 2005. Asymmetrical hysteresis in exchange-biased multilayers with out-of-plane applied fields, Phys. Reivew B. 71, 012417. doi:https://doi.org/10.1103/PhysRevB.71.012417.
\bibitem{Peck2011} M.. Peck, Y. Huh, R. Skomski, R. Zhang, P. Kharel, M.. Allison, D. Sellmyer, M. Langell, 2011. Magnetic properties of NiO and (Ni,Zn)O nanoclusters, Jounal Appl. Phys. 109, 07B518. doi:https://doi.org/10.1063/1.3556953
\bibitem{Meneses2010} C.T. Meneses, J.G.S. Duque, E. Biasi de, W.. Nunes, S. Sharma, M. Knobel, 2010. Competing interparticle interactions and surface anisotropy in NiO nanoparticles, J. Appl. Phys. 108, 013909. doi:https://doi.org/10.1063/1.3459890
\bibitem{Boukhari2019} J. Al Boukhari, L. Zeidan, A. Khalaf, R. Awad, Synthesis, characterization, optical and magnetic properties of pure and Mn, Fe and Zn doped NiO nanoparticles, Chem. Phys. 516 (2019) 116–124. doi:https://doi.org/10.1016/j.chemphys.2018.07.046
\bibitem{He2015} X. He, W. Zhong, Y. Du, 2015. Hexagonal CoO nanoparticles as studied by electron spin resonance, J. Appl. Phys. 117, 043905. doi:https://doi.org/10.1063/1.4906550
\bibitem{Carlin1935} R.L. Carlin, V.A. Duyneveldt, Magnetic Properties of Transition Metal compounds, Springer-Verlag, New york Heidelberg Berlin, 1935. doi:10.1 007/978-3-642-87392-8
\bibitem{Das2017} J. Das, D.K. Mishra, V. V Srinivasu, Spin canting and magnetism in nano-crystalline Zn$_{1-x}$Al$_x$O, J. Alloys Compd. 704 (2017) 237–244. doi:https://doi.org/10.1016/j.jallcom.2017.02.025
\bibitem{Golosovsky2005} I. V Golosovsky, I. Mirebeau, V. P. Sakhnenko, D. A. Kurdyukov, and Y. A. Kumzerov 2005. Evolution of the magnetic phase transition in MnO confined to channel type matrices: Neutron diffraction study, Phys. Rev. B. 72, 144409. doi:https://doi.org/10.1103/PhysRevB.72.144409
\bibitem{Gomonay2007} H. V Gomonay, V.M. Loktev, 2007. Shape-induced phenomena in finite-size antiferromagnets, Phyiscal Rev. B. 75, 174439. doi:https://doi.org/10.1103/PhysRevB.75.174439
\bibitem{Montes2014} R.N. Montes, P. Gorria, D. Martinez-Blanco, A.B. Fuertes, L. Fernandez, Barquin, J. Rodriguez Fernandez, I. de Pedro, M. Fdez-Gubieda, J. Alonso, L. Olivi, G. Aquilanti, J.A. Blanco, 2014. Interplay between microstructure and magnetism in NiO nanoparticles: breakdown of the antiferromagnetic order, Nanoscale. 6, 457. doi:10.1039/c3nr03961g
\bibitem{Thota2007} S. Thota, J. Kumar, Sol–gel synthesis and anomalous magnetic behaviour of NiO nanoparticles, J. Phys. Chem. Soilds. 68 (2007) 1951–1964. doi::10.1016/j.jpcs.2007.06.010
\bibitem{Mao2019} A. Mao, H.-Z. Xiang, Z.-G. Zhnag, K. Kuramoto, H. Yu, S. Ran, Solution combustion synthesis and magnetic property of rock-salt (Co$_{0.2}$Cu$_{0.2}$Mg$_{0.2}$Ni$_{0.2}$Zn$_{0.2}$)O high-entropy oxide nanocrystalline powder, J. Magn. Magn. Mater. 484 (2019) 245–252. doi:10.1016/j.jmmm.2019.04.023
\bibitem{Ravikumar2015} P. Ravikumar, B. Kisan, A. Perumal,2015. Enhanced room temperature ferromagnetism in antiferromagnetic NiO nanoparticles, AIP Adv. 5, 087116. doi:10.1063/1.4928426.
\bibitem{Joshi2018} D.C. Joshi, P. Pramanik, R.T. George, T. Sarkar, S. Thota, Nature of magnetic ordering in nanocomposites of Zn$_{1- p}$Ni$_p$O and NiO, Phys. E  Low-Dimensional Syst. Nanostructures. 103 (2018) 46–52. doi:10.1016/j.physe.2018.05.007.
\bibitem{Bracconi1983} P. Bracconi, Molecular-feild treatment of the high temperature susceptibility and Neel temperature of type II Antiferromagnetic solid-solutions xNiO-(1-x)CoO, J. Magn. Magn. Mater. 40 (1983) 37–47. doi:https://doi.org/10.1016/0304-8853(83)90008-2
\bibitem{Funkenbusch1981} E.. Funkenbusch, B.. Cornilsen, Two-magnon Raman scattering in calcium doped NiO, Solid State Commun. 40 (1981) 707–710. doi:https://doi.org/10.1016/0038-1098(81)90624-4.
\bibitem{Rost2017} C.M. Rost, Z. Rak, D.W. Brenner, J.-P. Maria, Local structure of the Mg$_x$Ni$_x$Co$_x$Cu$_x$Zn$_x$O (x=0.2) entropy-stabilized oxide: An EXAFS study, J. Am. Ceram. Soc. 100 (2017) 2732–2738. doi:10.1111/jace.14756.
\bibitem{Goodenough1955} J.B. Goodenough, A.L. Loeb, Theory of ionic ordering, crystal distortion, and magnetic exchange due to covalent forces in spinels, Phys. Rev. 98 (1955) 391–408. doi:https://doi.org/10.1103/PhysRev.98.391.
\end{thebibliography}

\end{document}